\documentstyle[12pt,titlepage,epsf]{article}
\def\_{\rule{.3em}{.15ex}} 
\def\bbuildrel#1_#2^#3{\mathrel{\mathop{\kern 0pt#1}\limits_{#2}^{#3}}}
\newcommand{\scb}{\scriptstyle}
\newcommand{\scs}{\scriptscriptstyle}
\newcommand{\be}{\begin{equation}}
\newcommand{\ee}{\end{equation}}
\newcommand{\bea}{\begin{eqnarray}}
\newcommand{\eea}{\end{eqnarray}}
\newcommand{\f}{\frac}

\newcommand{\e}{\epsilon}
\newcommand{\newton}[2]{ \left( \begin{array}{c} 
{\scb #1} \\ {\scb #2} \end{array} \right) }
\begin{document}
\begin{titlepage}

 \begin{flushright}
  {\bf MPI/PhT/97-45\\
       TTP97-43$^{^{\dagger}}$ \\
       ZU-TH-16/97\\
       TUM-HEP-284/97\\       
       IFT-11/97\\
      hep-ph/9711266\\
}
 \end{flushright}

 \begin{center}
  \vspace{0.6in}

{\Large \bf
Beta functions and anomalous dimensions\\ up to three loops}
\vspace{2cm} \\
\setlength {\baselineskip}{0.2in}

{\large  Konstantin Chetyrkin$^{^{1,\star}}$, 
         Miko{\l}aj Misiak$^{^{2}}$
         and Manfred M{\"u}nz$^{^{3}}$}\\

\vspace{0.2in}
$^{^{1}}${\it Institut f{\"u}r Theoretische Teilchenphysik, Universit{\"a}t Karlsruhe,\\
		   D-76128 Karlsruhe, Germany}

\vspace{0.2in}
$^{^{2}}${\it Institute of Theoretical Physics, Warsaw University,\\
		 PL-00-681 Warsaw, Poland}

\vspace{0.2in}
$^{^{3}}${\it Physik Department, Technische Universit{\"a}t M{\"u}nchen,\\
                         D-85748 Garching, Germany}

\vspace{2cm} 
{\bf Abstract \\}   
\end{center} 
\setlength{\baselineskip}{0.3in} 

	We derive an algorithm for automatic calculation of
perturbative $\beta$-functions and anomalous dimensions in any local
quantum field theory with canonical kinetic terms. The infrared
rearrangement is performed by introducing a common mass parameter in
all the propagator denominators. We provide a set of explicit formulae
for all the necessary scalar integrals up to three loops.

\vspace{2cm}

\setlength {\baselineskip}{0.2in}
\noindent \underline{\hspace{2in}}\\ 
\noindent 
$^{^{\dagger}}${\footnotesize The complete postscript file of this
  preprint, including figures, is available via anonymous ftp at
  www-ttp.physik.uni-karlsruhe.de (129.13.102.139) as /ttp97-43/ttp97-43.ps
  or via www at http://www-ttp.physik.uni-karlsruhe.de/cgi-bin/preprints.}

\noindent
$^\star$ {\footnotesize 
Permanent address: Institute of Nuclear Research,
Russian Academy of Sciences, Moscow 117312, Russia.}

\end{titlepage} 

\setlength {\baselineskip}{0.3in}

\noindent {\bf 1. Introduction}

	Renormalization group equations are a fundamental tool in
modern quantum field theory. In phenomenological applications, their
evaluation with sufficient accuracy often requires finding multiloop
contributions to $\beta$-functions and anomalous dimensions. In the
present paper, we describe a simple algorithm for calculating these
quantities in the framework of dimensional regularization and the $MS$
(or $\overline{MS}$) scheme. 

	In a mass-independent renormalization scheme,
$\beta$-functions and anomalous dimensions are simply related to
coefficients at counterterms which renormalize ultraviolet
divergences. A remarkable feature of the $MS$-scheme is the fact that
in its framework all the UV counter-terms are polynomial {\em both} in
momenta {\em and} in masses \cite{Collins75}.\footnote{
In any meaningful renormalization prescription, counterterms are
polynomial in external momenta, but not necessarily in masses.}
Consequently, a certain expansion in external momenta and masses can
be performed before integration over loop momenta, which radically
simplifies the integrals one needs to calculate.

	The main difficulty in this procedure is appearance of
spurious infrared divergences. The classical method of avoiding them
is called "infrared rearrangement" \cite{V80,CKT80}. It amounts to
adding artificial masses or external momenta in certain lines of a
given Feynman diagram before the expansion in masses and true external
momenta is made. The artificial external momenta have to be introduced
in such a way that all spurious infrared divergences are removed, and
the resulting Feynman integrals are calculable. Satisfying these two
requirements is rather cumbersome in practical multiloop calculations.
In addition, the condition that the IR divergences do not appear
restricts considerably the power of the approach, since for
complicated diagrams this requirement prevents one from reducing a
given Feynman diagram to a simpler one.

	The latter problem was completely solved with elaborating a
special technique of subtraction of IR divergences --- the
$R^*$-operation \cite{me84}. This method allows one to express (though
in a rather involved way) the UV counterterm of every (h+1)-loop
Feynman integral in terms of divergent and finite parts of some
properly constructed h-loop massless propagators. Unfortunately, in
practical applications, the use of the $R^*$-operation requires either
many manipulations with individual diagrams or resolving a lot of
non-trivial problem-dependent combinatorics (see, e.g. \cite{gssq,gvvq}).
	
	In our approach, the infrared rearrangement is performed by
introducing an artificial mass rather than an artificial external
momentum. A single mass parameter is added to each denominator of a
propagator in each Feynman diagram. Consequently, no spurious IR
divergences can appear. Next, an expansion in all the particle masses
(except, of course, the auxiliary one) and external momenta is
performed. The integrals one is left with have relatively simple form:
They are completely massive tadpoles, i.e. Feynman integrals without
external momenta and with only a single mass inserted in all the
propagators. As a result, the problem of evaluating h-loop UV
counterterms eventually reduces to a computation of divergent parts of
h-loop completely massive tadpoles.

	At two loops, simple formulae for such Feynman integrals have
been known since long ago \cite{BV84}. However, no explicit formulae
for three-loop massive tadpoles have been published so far. The
available recursion algorithms \cite{Broad82,Avdeev95} based on the
integration by parts method \cite{T81,CT81} are quite involved.

	The basic idea of our algorithm is to determine the pole part
of a massive tadpole by expanding a properly chosen two-loop
sub-integral with respect to its large external momentum being a loop
momentum in the initial three-loop integral. Eventually, we have been
able to construct relatively simple {\em explicit} formulae for all
the necessary three-loop scalar integrals. 

	The algorithm described in the present paper was used at the
two loop \cite{MM95} and three loop \cite{CMM96} levels for
calculating QCD anomalous dimensions of effective operators mediating
$B \to X_s \gamma$ decay.

	In principle, our method is applicable at the four-loop level,
too. In this case, the problem eventually amounts to expanding a
three-loop massive sub-integral of the propagator type with respect to
its large external momentum. The algorithm for calculation of such
three-loop integrals has been known since long ago (for a review see
\cite{smirnov95}) and its computer algebra implementation has been
recently achieved \cite{LME}.

	Very recently, an alternative algorithm was developed by van
Ritbergen, Vermaseren and Larin, and applied for evaluating four-loop
contributions to the QCD $\beta$ function and quark mass anomalous
dimension \cite{RVL97}. Their approach amounts to using an identical
to ours version of the IR rearrangement which reduces the calculation
of UV renormalization constants to calculation of massive tadpoles.
The difference appears at the stage of tadpole evaluation. The authors
of \cite{RVL97} have succeeded in creating "special routines (...) to
efficiently evaluate 4-loop massive bubble integrals up to pole parts
in $\e$ and correspondingly of the 3-loop massive bubbles to finite
parts." Eventually, all the diagrams have been reduced to two master
ones.

	Our paper is organized as follows: In the next section, we
give general arguments which justify the use of an artificial mass
parameter as an infrared regulator in all the propagators, including
propagators of massless gauge bosons. This is allowed so long as we
are interested only in the UV-divergent parts of regularized Green's
functions (with all UV {\em sub}divergences being pre-subtracted). In
section 3, we describe our algorithm for evaluating scalar integrals
up to three loops. In section 4, we present some more details
concerning calculation of nontrivial three-loop integrals. Section 5
contains two examples of relations between renormalization constants
and $\beta$ functions or anomalous dimensions up to three loops.
Appendix~A is devoted to reduction of tensor integrals to scalar ones.
Appendix B summarizes expressions for ``trivial'' integrals, i.e. the
ones which reduce to products of lower-loop integrals. Appendix C
describes the expansion of one-loop self-energy integrals at large
external momentum, which constitutes an essential element in
calculating nontrivial three-loop integrals. Finally, appendix D
contains a useful relation between tensor and scalar one-loop
integrals in different numbers of dimensions.

\ \\
{\bf 2. Decomposition of propagators}

	The starting point of our procedure is a certain exact
decomposition of propagators. For a scalar propagator
belonging to a given Feynman diagram, it has the following form:
\be \label{dec1}
\f{1}{(q+p)^2 - M^2} = \f{1}{q^2-m^2} 
+ \f{M^2 - p^2 - 2qp - m^2}{q^2-m^2} \f{1}{(q+p)^2 - M^2}.
\ee

Here, $p$ is a linear combination of external momenta in the
considered diagram, $q$ stands for a linear combination of loop
momenta, and $M$ denotes the mass of the particle.  The
artificial mass parameter $m$ is introduced to regularize
spurious infrared divergences.  It is the same in all the
propagators and all the diagrams.

	The contribution of the considered propagator to the
overall degree of divergence of a diagram is $\Delta \omega =-2$.
The decomposition has been performed in such a way that the
first simple term in the r.h.s. of eqn.~(\ref{dec1}) gives
$\Delta \omega =-2$, while the second, more complicated term gives
$\Delta \omega =-3$. Moreover, the very last term in eqn.~(\ref{dec1})
has the same form as the original propagator. Thus, we can
decompose it in an identical way. Doing so several times,
we decompose the original propagator into a sum of terms with
very simple denominators (depending only on loop momenta and the
mass parameter $m$), and a more complicated term whose contribution
to the overall degree of divergence is arbitrarily low negative.
For instance, after three steps of decomposition, the exact
expression for the original propagator reads
\bea \label{dec3}
\f{1}{(q+p)^2 - M^2} &=& 
    \f{1}{q^2-m^2} + \f{M^2-p^2-2qp}{(q^2-m^2)^2} + 
    \f{(M^2-p^2-2qp)^2}{(q^2-m^2)^3} \nonumber \\
&& - \f{m^2}{(q^2-m^2)^2} + \f{m^4 - 2m^2 (M^2 - p^2 - 2qp)}{(q^2-m^2)^3}
\nonumber \\
&& + \f{(M^2-p^2-2qp-m^2)^3}{(q^2-m^2)^3 [(q+p)^2 - M^2]}.
\eea
Here, the last term gives $\Delta \omega =-5$ contribution to the
overall degree of divergence of a diagram.

	In the following, we shall assume that the theory we consider
is given by an (effective) lagrangian which does not contain
non-negligible operators of arbitrarily high dimension, i.e. we assume
that dimensionality of our operators is bounded from above. In such a
case, any particular Green's function has a certain maximal degree of
divergence. Consequently, we can always perform so many steps in the
propagator decomposition, that the overall degree of divergence of any
diagram in this Green's function would become negative if any of its
propagators was replaced by the last term in the decomposition. We are
then allowed to drop the last term in each propagator decomposition.
It does not affect the UV-divergent part of the Green's function ({\em
after} subtraction of subdivergences).

	It is important to note that each term in the propagator
decomposition satisfies the criteria a full propagator should satisfy
in the proof of Weinberg's theorem \cite{W60}. This allows to apply
degree-of-divergence arguments for diagrams where propagators are
replaced by particular terms in their decomposition.

	A further simplification can be achieved by noticing that
terms containing $m^2$ in the numerators (like the second line in the
r.h.s. of eqn.~(\ref{dec3})) contribute only to such UV-divergent
terms which are proportional to $m^2$. These terms are local after
subtraction of subdivergences. They must precisely cancel similar
terms originating from integrals with no $m^2$ in propagator
numerators. No dependence on $m^2$ can remain after performing the
whole calculation, because the propagators are decomposed exactly,
i.e. they are actually independent of $m^2$. This observation allows
to avoid calculating integrals with $m^2$ in propagator numerators.
Instead of calculating them, one can just replace them by local
counterterms proportional to $m^2$ which cancel the corresponding
(sub)divergences in integrals with no $m^2$ in propagator numerators.
In effect, the practical calculation is made only with propagators
replaced by such terms in the decomposition which contain no $m^2$ in
the numerators (like the first line in the r.h.s. of
eqn.~(\ref{dec3})). Nevertheless, the final results for the divergent
parts of Green's functions are precisely the same as if the full
propagators were used ({\em after} subtraction of subdivergences).

	In our earlier paper \cite{MM95}, we gave somewhat
different arguments for using a single mass parameter as an
infrared regulator in calculating $\beta$ functions and
anomalous dimensions. The present considerations might be more
convincing, because the propagator decomposition we discuss here
is exact. Thus, $m^2$ can be kept arbitrary all the time.  One
does not need to consider the $m^2 \to 0$ limit and worry about
its commutativity with Feynman integration.

	For particles with spin other than zero, the decomposition is
applied only to denominators of their propagators, provided they are
the same as in the scalar propagator. Our algorithm is not applicable
in theories where kinetic terms differ from the canonical ones, as
e.g. in the Heavy Quark Effective Theory \cite{N94}.

	As we have explained, one does not need to calculate Feynman
integrals containing $m^2$ in propagator numerators, so long as extra
counterterms proportional to $m^2$ are introduced.  Such counterterms
may not preserve symmetries of the theory.  Fortunately, the number of
these counterterms is usually rather small, because their dimension
must be at least twice smaller than the maximal dimension of operators
in the considered (effective) lagrangian. For instance, the QCD
lagrangian is built out of operators of dimension less or equal
4. There is only a single possible gauge-noninvariant counterterm of
dimension 2. It reads
\be \label{gluon.mass}   
\f{1}{2} Z_x m^2 G_{\mu}^a G^{a\;\mu},
\ee
i.e. it looks like a "gluon mass" counterterm.\footnote{
The ``ghost mass'' counterterm does not arise in the Feynman--'t~Hooft
gauge, due to the structure of the ghost-gluon vertex.}
At one loop, we find (using the Feynman--'t~Hooft gauge and the $MS$
scheme in $D = 4 - 2 \e$ dimensions)
\be \label{zx}
Z_x = -\f{g^2}{16 \pi^2 \e} (N + 2f),
\ee
where $N$ is the number of colors and $f$ is the number of active
flavors. This counterterm cancels gauge-noninvariant pieces of integrals
with no $m^2$ in propagator numerators.\\

	After dropping the last term in the propagator
decomposition, the Feynman integrands one is left with depend
only polynomially on particle masses and external momenta.
These quantities can be factorized out. It remains to calculate
integrals depending only on loop momenta and the artificial mass
parameter $m^2$. At one loop, the generic integral reads
\be \label{1loop.tens.int}
\int \f{d^{\scs D}q}{(2\pi)^D} \f{q_{\mu_1} ... q_{\mu_k}}{(q^2-m^2)^n}.
\ee
Integrals arising at more loops are slightly more complicated, because
they involve several loop momenta. Nevertheless, reducing any such
integral to scalar integrals can be easily performed by contracting it
with various products of metric tensors and solving the resulting
system of linear equations. We have written a Mathematica \cite{MMa}
code which performs such a reduction up to three loops, for an
arbitrary number of free Lorentz indices.  Some elements of this
procedure are outlined in appendix A.

\begin{figure}[h] 
\centerline{
\epsfysize = 3cm
\epsffile{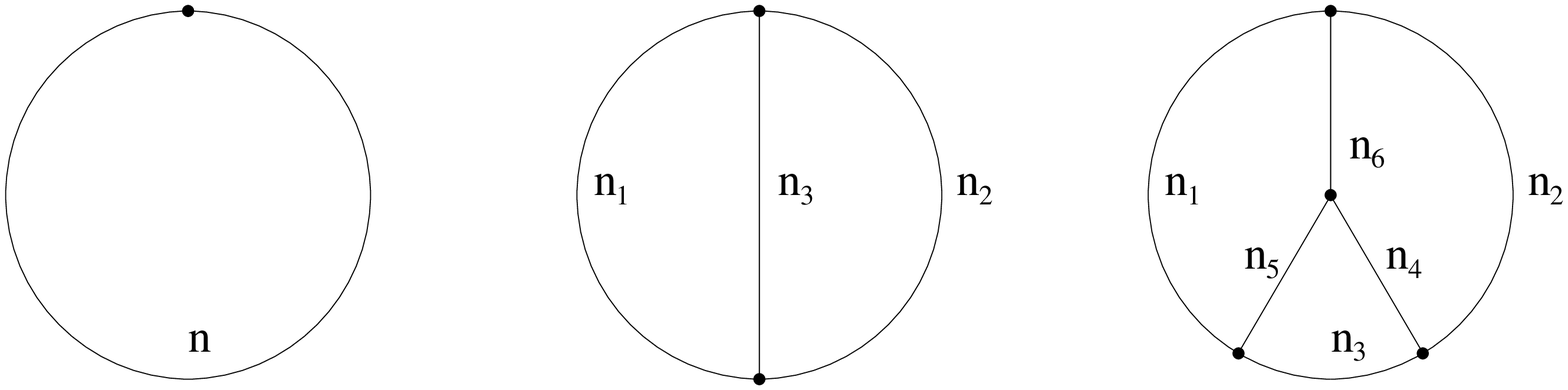}}
\caption{Graphical representation of the integrals given in 
          eqns.~(\ref{int1})--(\ref{int3}) \label{vac.diag}}
\end{figure}

	After the reduction of tensor integrals is performed, one is
left with relatively small number of scalar integrals to calculate. It
is convenient to use the euclidean metric in discussing their
evaluation. The euclidean integrals arising at one, two and three
loops are respectively as follows
\bea 
I^{(1)}_n &=& m^{-{\scs D}+2n}\; \pi^{-\f{D}{2}} \int d^{\scs D} q \f{1}{[q^2 + m^2]^n},
\label{int1}\\
I^{(2)}_{n_1 n_2 n_3} &=& m^{-2{\scs D}+2\Sigma n_i}\; \pi^{-{\scs D}} 
\int d^{\scs D} q_1 \; d^{\scs D} q_2 
\f{1}{[q_1^2 + m^2]^{n_1} [q_2^2 + m^2]^{n_2} [(q_1-q_2)^2 + m^2]^{n_3}},\;\;\;\;
\label{int2}\\
I^{(3)}_{n_1 n_2 n_3 n_4 n_5 n_6} &=&  m^{-3{\scs D}+2\Sigma n_i}\; \pi^{-\f{3D}{2}} 
\int d^{\scs D} q_1 \; d^{\scs D} q_2 \; d^{\scs D} q_3 \times 
\nonumber 
\eea
\be
\times
\f{1}{[q_1^2 + m^2]^{n_1} [q_2^2 + m^2]^{n_2} [q_3^2 + m^2]^{n_3}
[(q_2-q_3)^2 + m^2]^{n_4} [(q_3-q_1)^2 + m^2]^{n_5} [(q_1-q_2)^2 + m^2]^{n_6}}.
\label{int3}
\ee
The chosen normalization makes them dimensionless. The integrals can
be represented by scalar vacuum diagrams displayed in
fig.~\ref{vac.diag} with propagators raised to arbitrary integer
powers $n_i$. The algorithm for their evaluation is described in the
next section.

\ \\
{\bf 3. The algorithm for evaluation of scalar integrals.}

	In this section, we assume we are interested in evaluating
three-loop $\beta$-functions or anomalous dimensions.  We use the MS
scheme with $D=4-2\e$ dimensions. We need to be able to evaluate
$I^{(3)}_{n_1 n_2 n_3 n_4 n_5 n_6}$ up to ${\cal O}(\f{1}{\e})$, 
$I^{(2)}_{n_1 n_2 n_3}$ up to ${\cal O}(1)$ and 
$I^{(1)}_n$ up to ${\cal O}(\e)$. 
The latter integral is known exactly from textbooks \cite{C84}
\be \label{1loop.scal.int}
I^{(1)}_n = \f{\Gamma(n-\f{D}{2})}{\Gamma(n)}.
\ee
The two-loop integral $I^{(2)}_{n_1 n_2 n_3}$ is totally symmetric
under permutations of its indices. It reduces to a product of one-loop
integrals when at least one of the indices is nonpositive (see
appendix B for explicit formulae). On the other hand, when all the
indices are positive, it can be found from the following relations
\cite{DT93}:
\bea \label{recursion}
I^{(2)}_{(n_1+1) n_2 n_3} = \f{1}{3 n_1} \{ (3 n_1 - D) I^{(2)}_{n_1 n_2 n_3}
&+& n_2 [ I^{(2)}_{(n_1-1) (n_2+1) n_3} - I^{(2)}_{n_1 (n_2+1) (n_3-1)}]
\nonumber \\
&+& n_3 [ I^{(2)}_{(n_1-1) n_2 (n_3+1)} - I^{(2)}_{n_1 (n_2-1) (n_3+1)}] \}
\eea
and 
\be \label{i111}
I^{(2)}_{111} = \f{\Gamma(1+\e)^2}{(1-\e)(1-2\e)} 
[ -\f{3}{2\e^2} + \f{27}{2} S_2 ] + {\cal O}(\e),
\ee
where
\be \label{S_2}
S_2 = -\f{4}{9 \sqrt{3}} \int_0^{\pi/3} dx \ln ( 2 \sin \f{x}{2} ) 
\simeq 0.2604341.
\ee
The recursion relation (\ref{recursion}) holds for $n_i \geq 1$.  It
can be derived with use of integration by parts. The sum of indices in
the integral on its l.h.s. is bigger than the sum of indices in each
of the integrals on its r.h.s.. Thus, the recursion can be programmed
into a computer algebra code just as it stands. Two-loop integrals one
usually encounters in practice are then found within a fraction of a
second.

\begin{figure}[h] 
\centerline{
\epsfysize = 3cm
\epsffile{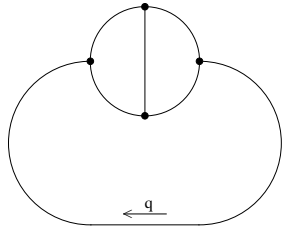}}
\caption{Graphical representation of the integral $I^{(3)}_{111111}$. 
It is equivalent to the last diagram in fig.~\ref{vac.diag}
with $n_1 = ... = n_6 = 1$. \label{merc.egg}}
\end{figure}

	Let us now turn to the three-loop integrals 
$I^{(3)}_{n_1 n_2 n_3 n_4 n_5 n_6}$. Here, we are interested in
calculating only UV-divergent parts of them. It is instructive
to subsequently consider three cases:
\begin{itemize}
\item{A:~} All the indices $n_1,...,n_6$ are positive.

\item{B:~} At least one of the indices is equal to zero.

\item{C:~} None of the indices vanishes, but some of them are
negative.
\end{itemize}

\noindent
{\bf Case A:~} From among all the three-loop integrals with six
positive indices, only $I^{(3)}_{111111}$ is UV-divergent.  Integrals
with larger positive indices have negative degree of divergence and no
subdivergences.

\begin{figure}[h] 
\centerline{
\epsfysize = 3cm
\epsffile{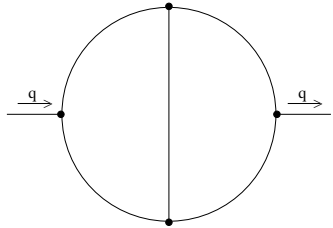}}
\caption{Graphical representation of the 
           integral $J^{(2)}_{11111}(q^2,m^2)$. \label{egg}}
\end{figure}

	In order to calculate the UV-divergent part of
$I^{(3)}_{111111}$ we write this integral as follows (see
fig.~\ref{merc.egg}):
\be \label{i111111}
I^{(3)}_{111111} \;\; = \;\; m^{-{\scs D}+2}\; \pi^{-\f{D}{2}} \int d^{\scs D} q
\f{1}{[q^2 + m^2]} J^{(2)}_{11111}(q^2,m^2),
\ee
where
\bea \label{j11111}
J^{(2)}_{11111}(q^2,m^2) \;\; = \;\; m^{-2D+10}\; \pi^{-{\scs D}} \times \hspace{10cm}
\nonumber 
\eea
\be
\times \int d^{\scs D} q_1 d^{\scs D} q_2
\f{1}{[q_1^2 + m^2] [q_2^2 + m^2]
[(q_1-q_2)^2 + m^2] [(q-q_1)^2 + m^2] [(q-q_2)^2 + m^2]}. 
\ee
The latter integral is just the usual two-loop contribution to the
wave function renormalization in the "$\lambda \phi^3$" theory. We
show it in fig.~\ref{egg}. It is a finite diagram, because it has
negative degree of divergence and no subdivergences. Consequently, a
finite-volume integration in eqn.~(\ref{i111111}) cannot give a $1/\e$
pole. Such a pole can only arise from integration over large $q^2$ in
eqn.~(\ref{i111111}). Therefore, knowing the behavior of
$J^{(2)}_{11111}(q^2,m^2)$ at large $q^2$ is enough to find the
UV-divergent part of $I^{(3)}_{111111}$.

	The two-loop integral $J^{(2)}_{11111}(q^2,m^2)$ has the following 
expansion at large $q^2$ \cite{CKT80}:
\be \label{exp.egg}
J^{(2)}_{11111}(q^2,m^2) 
\bbuildrel{=\!=\!=\!=}_{\scs q^2 >> m^2}^{}
\left( \f{m^2}{q^2} \right)^{5-D} \left[ 
6 \zeta(3) + {\cal O}\left( \f{m^2}{q^2} \right) + {\cal O}(\e) \right].
\ee
Inserting this result into eqn.~(\ref{i111111}) and introducing an
infrared cutoff $\Lambda$ one finds
\bea \label{result.111111}
I^{(3)}_{111111} &=& \f{1}{\Gamma(\f{D}{2})} \int_{\Lambda}^\infty d (q^2) 
\left( \f{q^2}{m^2} \right)^{\f{D}{2}-1}
\f{1}{[q^2 + m^2]} J^{(2)}_{11111}(q^2,m^2) 
\; + \; \mbox{(finite terms)}
\nonumber \\
&=& \left( \f{m}{\Lambda} \right)^{6\e} \f{2 \zeta(3)}{\e}
\; + \; \mbox{(finite terms)}
\nonumber \\
&=& \f{2 \zeta(3)}{\e} \; + \; \mbox{(finite terms)}.
\eea

	The way we have found the UV-divergent part of
$I^{(3)}_{111111}$ shows the basic idea for calculating all the
nontrivial three-loop integrals in the cases B and C.  The
UV-divergent parts of these integrals can be found by choosing some
two-loop subdiagrams of the last graph in fig.~\ref{vac.diag} and
considering their behavior at large external momenta. If the
considered two-loop subdiagram is finite, the calculation proceeds
analogously to the case of $I^{(3)}_{111111}$. If it is divergent, a
subtraction of the UV divergence needs to be performed. We describe
this in more detail below.

\ \\
{\bf Case B:~} Now, we consider the case when at least one of the
indices of $I^{(3)}_{n_1 n_2 n_3 n_4 n_5 n_6}$ is equal to zero.
Without loss of generality, we can assume that the vanishing
index is $n_6$. This is because of the tetrahedron symmetry: The
last diagram in fig.~\ref{vac.diag} has the topology of a
tetrahedron.  Symmetries of a tetrahedron can be described as
certain permutations of its edges. Such permutations of the
indices $(n_1,...,n_6)$ leave our integral invariant.

\begin{figure}[h] 
\centerline{
\epsfysize = 3cm
\epsffile{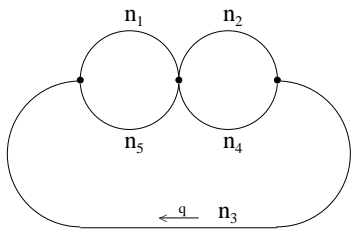}}
\caption{Graphical representation of the three-loop integral
         with one nonpositive index \label{double.bubble}}
\end{figure}

	When $n_6 = 0$, the integral can be written as
\be \label{eq.double.bubble}
I^{(3)}_{n_1 n_2 n_3 n_4 n_5 0} =
m^{-{\scs D}+2 n_3}\; \pi^{-\f{D}{2}} \int d^{\scs D} q \f{1}{[q^2 + m^2]^{n_3}}
G_2\left(n_1,n_5,\f{m^2}{q^2}\right) 
G_2\left(n_2,n_4,\f{m^2}{q^2}\right),
\ee
where the one-loop integral $G_2$ is given by
\be \label{G2.def}
G_2\left(k_1, k_2, \f{m^2}{q^2}\right) = 
m^{-{\scs D} + 2 k_1 + 2 k_2}\; \pi^{-\f{D}{2}}
\int d^{\scs D} p \f{1}{[p^2 + m^2]^{k_1} [(p-q)^2 + m^2]^{k_2}}.
\ee
The diagram corresponding to eqn.~(\ref{eq.double.bubble}) is
shown in fig.~\ref{double.bubble}.  

	Calculation of the integral (\ref{eq.double.bubble}) depends
on what values are taken by indices of the one-loop sub-integrals
denoted by $G_2$.\footnote{
However, it does not make any qualitative difference whether the index
$n_3$ in the final integration is positive or not.}
One can distinguish three situations:

\begin{itemize}

\item{B.1:~} When at least one of these indices ($n_1,n_2,n_4$ or
$n_5$) is nonpositive, then the three-loop integral reduces to a
product of one- and two-loop tensor integrals. The latter integrals
can be easily reduced to scalar integrals which we are already able to
calculate. Final formulae for such three-loop integrals are given in
appendix~B.

\item{B.2:~} When $n_1, n_2, n_4, n_5 > 0$ and both integrals
$G_2$ are convergent (i.e. $n_1 + n_5 > 2$ and $n_2 + n_4 > 2$),
the calculation proceeds analogously to the case of
$I^{(3)}_{111111}$.  We expand the convergent integrals $G_2$ at
large $q^2$
\be \label{G2.exp}
G_2\left(k_1,k_2,\f{m^2}{q^2}\right) 
\bbuildrel{=\!=\!=\!=}_{\scs q^2 \to \infty}^{} \sum_{r=0}^{\infty} 
\left[ \left( \f{m^2}{q^2} \right)^r      a(k_1,k_2,r)
   +   \left( \f{m^2}{q^2} \right)^{r+\e} b(k_1,k_2,r) \right].
\ee
Only a few lowest terms in this expansion affect the pole part
of the considered three-loop integral. Explicit expressions for
$a(k_1,k_2,r)$ and $b(k_1,k_2,r)$ are given in appendix~C.

\item{B.3:~} When $n_1, n_2, n_4, n_5 > 0$ but one or both
integrals $G_2$ are divergent (i.e. $n_1 = n_5 = 1$ and/or $n_2
= n_4 = 1$), we need to split the integral $G_2$ into its pole
and convergent parts
\be \label{G2.ren}
G_2\left(k_1,k_2,\f{m^2}{q^2}\right) =
\f{1}{\e} \delta_{1k_1} \delta_{1k_2} + 
G_2^{ren}\left(k_1,k_2,\f{m^2}{q^2}\right).
\ee

	Inside the three-loop integral, the pole part of $G_2$
is multiplied by a two-loop integral. Thus, we already know how
to calculate its contribution to the UV-divergence of the
three-loop integral. On the other hand, the "renormalized" part
of $G_2$ can be treated analogously to the case B.2, i.e. in
the same way the whole $G_2$ was treated when it was convergent.
Expansion of the "renormalized" integral $G_2^{ren}$ at large
$q^2$ is identical as in eqn.~(\ref{G2.exp}), except for that
$a(k_1,k_2,r)$ is replaced by
\be \label{a.ren}
a^{ren}(k_1, k_2, r) = 
a(k_1, k_2, r) - \f{1}{\e} \delta_{1k_1} \delta_{1k_2} \delta_{0r}.
\ee
One should not naively expect that $a^{ren}(k_1,k_2,r)$ contain no
poles in $\e$. Actually, they do contain simple poles which cancel
with the poles of $b(k_1,k_2,r)$ in the expression for $G_2^{ren}$.
\end{itemize}

\ \\
{\bf Case C:~} Now, we consider integrals with some negative
indices but with none of them equal to zero. Using the
tetrahedron symmetry, we can assume without loss of generality
that one of the negative indices is $n_6$.  Then, we consider
two distinct situations:

\begin{itemize}
\item{C.1:~} When any of the indices $n_1,n_2,n_4$ or $n_5$ is
negative, the three-loop integral reduces to products of one-
and two-loop integrals, similarly to the case when $n_6=0$. The
explicit formulae given in appendix~B apply both when $n_6$
vanishes and when it is negative.

\item{C.2:~} When all the remaining indices are positive or the only
other negative index is $n_3$, we can still represent the considered
three-loop integral by the diagram shown in
fig.~\ref{double.bubble}. However, instead of the scalar one-loop
integrals $G_2$, we encounter tensor one-loop integrals.  This does not
lead to any real difficulty, because we are able to reduce tensor
integrals to scalar ones. Nevertheless, the amount of necessary
algebra can be drastically reduced when one makes use of certain
tensor identities discussed in appendix~D.
\end{itemize}

	In the above considerations, we have described a complete
algorithm for calculating pole parts of the integrals defined in
eqns.~(\ref{int1})--(\ref{int3}). However, obtaining final formulae
for nontrivial three-loop integrals in the cases B.2, B.3 and C.2
requires discussing a few more subtle points in their evaluation.  This
is what the next section is devoted to.

\ \\
{\bf 4. More on nontrivial three-loop integrals.}

	Let us first derive our final expression for the three-loop
integrals in the cases B.2 and B.3. In both these cases, the
considered three-loop integral can be written as
\bea \label{longest}
I^{(3)}_{n_1 n_2 n_3 n_4 n_5 0} &=&
m^{-{\scs D}+2 n_3}\; \pi^{-\f{D}{2}} \int d^{\scs D} q 
\f{1}{[q^2 + m^2]^{n_3}}
G^{ren}_2\left(n_1,n_5,\f{m^2}{q^2}\right) 
G^{ren}_2\left(n_2,n_4,\f{m^2}{q^2}\right)
\nonumber \\ 
&& + \f{1}{\e} \delta_{1n_1} \delta_{1n_5} I^{(2)}_{n_2 n_4 n_3}
   + \f{1}{\e} \delta_{1n_2} \delta_{1n_4} I^{(2)}_{n_1 n_5 n_3}
   - \f{1}{\e^2} \delta_{1n_1} \delta_{1n_2} \delta_{1n_4} \delta_{1n_5} 
     I^{(1)}_{n_3} =
\nonumber \\ 
&& \nonumber \\
&=& \mbox{(finite terms)} + \f{1}{\Gamma(\f{D}{2})} \int_0^{\f{m^2}{\Lambda^2}}
d\left(\f{m^2}{q^2}\right) \left( 1 + \f{m^2}{q^2} \right)^{-n_3} 
\sum_{r_1,r_2=0}^{\infty} \left(\f{m^2}{q^2}\right)^{n_3+r_1+r_2-3} \times
\nonumber \\ 
&& \times \left\{ 
  a^{ren}(n_1,n_5,r_1) a^{ren}(n_2,n_4,r_2) \left(\f{m^2}{q^2}\right)^{\e}
+       b(n_1,n_5,r_1)       b(n_2,n_4,r_2) \left(\f{m^2}{q^2}\right)^{3\e} 
\right. \nonumber \\ && \hspace{1cm} \left.
+ \left[   a^{ren}(n_1,n_5,r_1) b(n_2,n_4,r_2) 
         + b(n_1,n_5,r_1) a^{ren}(n_2,n_4,r_2) \right] \left(\f{m^2}{q^2}\right)^{2\e}
\right\}
\nonumber \\ 
&& + \f{1}{\e} \delta_{1n_1} \delta_{1n_5} I^{(2)}_{n_2 n_4 n_3}
   + \f{1}{\e} \delta_{1n_2} \delta_{1n_4} I^{(2)}_{n_1 n_5 n_3}
   - \f{1}{\e^2} \delta_{1n_1} \delta_{1n_2} \delta_{1n_4} \delta_{1n_5} 
     I^{(1)}_{n_3}.
\eea
Similarly to eqn.~(\ref{result.111111}), an arbitrary infrared cutoff
$\Lambda$ has been introduced here. Assuming that $\Lambda^2 \ge m^2$,
we can expand in eqn.~(\ref{longest})
\be
\left( 1 + \f{m^2}{q^2} \right)^{-n_3} = \sum_{k=0}^{\infty} 
\f{(-n_3) (-n_3-1) ... (-n_3 - k + 1)}{k!} \left(\f{m^2}{q^2}\right)^k 
\equiv \sum_{k=0}^{\infty} \newton{-n_3}{k} \left(\f{m^2}{q^2}\right)^k. \hspace{2cm} 
\ee
After performing trivial integrations, we arrive at the following
result:
\bea \label{inf.sum}
I^{(3)}_{n_1 n_2 n_3 n_4 n_5 0} &=&
\mbox{(finite terms)} + \f{1}{\Gamma(\f{D}{2})} \sum_{r_1,r_2,k=0}^{\infty}
\newton{-n_3}{k} \left(\f{m^2}{\Lambda^2}\right)^{n_3 - 2 + r_1 + r_2 + k} \times
\nonumber \\
&& \times \left\{ 
  \f{a^{ren}(n_1,n_5,r_1) a^{ren}(n_2,n_4,r_2)}{n_3-2+r_1+r_2+k+\e}
						\left(\f{m^2}{\Lambda^2}\right)^{\e}
+ \f{      b(n_1,n_5,r_1)       b(n_2,n_4,r_2)}{n_3-2+r_1+r_2+k+3\e}
						\left(\f{m^2}{\Lambda^2}\right)^{3\e}
\right. \nonumber \\ && \hspace{1cm} \left.
+ \f{a^{ren}(n_1,n_5,r_1) b(n_2,n_4,r_2) + b(n_1,n_5,r_1) a^{ren}(n_2,n_4,r_2)
}{n_3-2+r_1+r_2+k+2\e} \left(\f{m^2}{\Lambda^2}\right)^{2\e} \right\}
\nonumber \\ 
&& + \f{1}{\e} \delta_{1n_1} \delta_{1n_5} I^{(2)}_{n_2 n_4 n_3}
   + \f{1}{\e} \delta_{1n_2} \delta_{1n_4} I^{(2)}_{n_1 n_5 n_3}
   - \f{1}{\e^2} \delta_{1n_1} \delta_{1n_2} \delta_{1n_4} \delta_{1n_5} 
     I^{(1)}_{n_3}.
\eea
The curly bracket in the above equation contains no $1/\e$ poles
unless $n_3 - 2 + r_1 + r_2 + k = 0$. Verifying this requires a short
calculation, because $a^{ren}(k_1,k_2,r)$ and $b(k_1,k_2,r)$ do
contain simple poles in $\e$. Thus, for $n_3 - 2 + r_1 + r_2 + k \neq
0$, one needs to expand the denominators to ${\cal O}(\e)$ and check
that the potential $1/\e$ contributions to $I^{(3)}$ ``miraculously''
sum up to zero, due to
\be
a^{ren}(k_1,k_2,r) + b(k_1,k_2,r) = {\cal O}(1).
\ee

	Our final expression for $I^{(3)}_{n_1 n_2 n_3 n_4 n_5 0}$ in
the cases B.2 and B.3 is thus given by a finite sum (from now on we
set $\Lambda = m$ for simplicity\footnote{
One could keep $\Lambda$ arbitrary and verify that the pole part of
$I^{(3)}$ is independent of this parameter. This can serve as a useful
cross-check against misprints in the explicit expressions for
$a(k_1,k_2,r)$ and $b(k_1,k_2,r)$ in appendix C.}
):
\bea \label{fin.sum}
I^{(3)}_{n_1 n_2 n_3 n_4 n_5 0} &=&
\mbox{(finite terms)} + \f{1}{\e \Gamma(2-\e)} 
\sum_{r_1=0}^{2-n_3} \;\; \sum_{r_2=0}^{2-n_3-r_1}    
\newton{-n_3}{2-n_3-r_1-r_2} \times
\nonumber \\
&& \times \left\{ a^{ren}(n_1,n_5,r_1) a^{ren}(n_2,n_4,r_2)
+ \f{1}{3} b(n_1,n_5,r_1) b(n_2,n_4,r_2) \right. 
\nonumber \\ && \hspace{1cm} 
\left. + \f{1}{2} \left[ a^{ren}(n_1,n_5,r_1) b(n_2,n_4,r_2) + 
            b(n_1,n_5,r_1) a^{ren}(n_2,n_4,r_2) \right] \right\}
\nonumber \\ 
&& + \f{1}{\e} \delta_{1n_1} \delta_{1n_5} I^{(2)}_{n_2 n_4 n_3}
   + \f{1}{\e} \delta_{1n_2} \delta_{1n_4} I^{(2)}_{n_1 n_5 n_3}
   - \f{1}{\e^2} \delta_{1n_1} \delta_{1n_2} \delta_{1n_4} \delta_{1n_5} 
     I^{(1)}_{n_3}.
\eea

\vspace*{0.2cm}

	Let us now turn to the most complicated case C.2. In this
case, the indices $n_1, n_2, n_4$ and $n_5$ are positive, while the
index $n_6$ is negative. Using the tensor identities given in the end
of appendix D, we express $[(q_1-q_2)^2+m^2]^{-n_6}$ in terms of
symmetric and traceless tensors
\bea
[(q_1-q_2)^2+m^2]^{-n_6} &=& \sum_{k=0}^{-n_6} \sum_{i=0}^k 
\sum_{\rho=0}^{[i/2]} \sum_{l_1=0}^{\rho} \;\; \sum_{l_2=0}^{k-i+\rho}
\newton{-n_6}{k} \newton{k}{i} \newton{\rho}{l_1} \newton{k-i+\rho}{l_2} 
\times \nonumber \eea \bea \times
\f{i! 2^i (-1)^{i+l_1+l_2} (m^2)^{l_1+l_2}
}{4^{\rho}\rho!(i-2\rho)! (i+2-2\rho-\e)_{\rho}} (q_1^2+m^2)^{-n_6-k+\rho-l_1}
(q_2^2+m^2)^{k-i+\rho-l_2} (q_1 \cdot q_2)^{(i-2\rho)},
\eea 
where $(x)_n$ denotes the Pochhammer symbol  
\be
(x)_n = x(x+1)(x+2)...(x+n-1) = \f{\Gamma(x+n)}{\Gamma(x)}.
\ee
Consequently, we can write
\bea \label{decomp.into.sym}
I^{(3)}_{n_1 n_2 n_3 n_4 n_5 n_6} &\bbuildrel{=\!=\!=}_{\scs n_6 < 0}^{}&
\sum_{k=0}^{-n_6} \sum_{i=0}^k \sum_{\rho=0}^{[i/2]} \sum_{l_1=0}^{\rho} 
\sum_{l_2=0}^{k-i+\rho} \newton{-n_6}{k} \newton{k}{i} \newton{\rho}{l_1} 
\newton{k-i+\rho}{l_2} \times \nonumber \\ && \times 
\f{i! 2^{i-2\rho} (-1)^{i+l_1+l_2}}{\rho!(i-2\rho)! (i+2-2\rho-\e)_{\rho}}
I^{(3)(i-2\rho)}_{(n_1+n_6+k-\rho+l_1)(n_2-k+i-\rho+l_2) n_3 n_4 n_5 0}
\eea 
where
\bea
I^{(3)(n)}_{n_1 n_2 n_3 n_4 n_5 0} &=&  m^{-3{\scs D} -2n + 2\Sigma n_i} \; 
\pi^{-\f{3D}{2}} \int d^{\scs D} q_1 \; d^{\scs D} q_2 \; d^{\scs D} q_3 \times 
\nonumber \\ 
&\times&
\f{(q_1 \cdot q_2)^{(n)}}{[q_1^2 + m^2]^{n_1} [q_2^2 + m^2]^{n_2} [q_3^2 + m^2]^{n_3}
[(q_2-q_3)^2 + m^2]^{n_4} [(q_3-q_1)^2 + m^2]^{n_5}}. \hspace{1cm}
\label{prefactor}
\eea

	The above integral is a generalization of $I^{(3)(0)}_{n_1 n_2
n_3 n_4 n_5 0} \equiv I^{(3)}_{n_1 n_2 n_3 n_4 n_5 0}$ considered in
the case B. When $n_1, n_2, n_4$ or $n_5$ is nonpositive,
$I^{(3)(n)}_{n_1 n_2 n_3 n_4 n_5 0}$ is equal to a linear combination
of reducible integrals considered in the cases B.1 and C.1.  The
explicit form of this linear combination is given in the end of
appendix B. On the other hand, when $n_1, n_2, n_4$ and $n_5$ are all
positive, the calculation of $I^{(3)(n)}_{n_1 n_2 n_3 n_4 n_5 0}$
proceeds analogously to the cases B.2 and B.3. However, instead of the
scalar integrals $G_2$, we encounter tensor one-loop integrals with
totally symmetric and traceless tensors in their numerators. Such
one-loop integrals are in one-to-one correspondence with scalar
one-loop integrals in larger number of dimensions. The appropriate
relation is given in appendix D.  Using this relation, one finds the
necessary generalization of eqn.~(\ref{eq.double.bubble})
\bea
I^{(3)(n)}_{n_1 n_2 n_3 n_4 n_5 0} &=& (n_4)_n (n_5)_n
m^{-4+2\e-2n+2 n_3}\; \pi^{-2+\e} \int d^{4-2\e} q 
\f{(q \cdot q)^{(n)}}{[q^2 + m^2]^{n_3}} 
\times \nonumber \\ &\times&
G_2\left.\left(n_1,n_5+n,\f{m^2}{q^2}\right)\right|_{D=4+2n-2\e} 
G_2\left.\left(n_2,n_4+n,\f{m^2}{q^2}\right)\right|_{D=4+2n-2\e},
\eea
where $(q \cdot q)^{(n)}$ can be expressed back in terms of $q^2$
\be
(q \cdot q)^{(n)} = \f{(2-2\e)_n}{2^n(1-\e)_n} (q^2)^n.
\ee

	Similarly to eqn.~(\ref{G2.ren}), we split the
higher-dimensional $G_2$ into its pole and convergent parts
\be
G_2\left.\left(k_1,k_2+n,\f{m^2}{q^2}\right)\right|_{D=4+2n-2\e} =
\f{1}{\e(n+1)!} \delta_{1k_1} \delta_{1k_2} \;\; + \;\;
G_2^{ren}\left.\left(k_1,k_2+n,\f{m^2}{q^2}\right)\right|_{D=4+2n-2\e}.
\ee
Next, we expand the convergent part at large $q^2$, as in
eqn.~(\ref{G2.exp})
\bea
G_2^{ren}\left.\left(k_1,k_2+n,\f{m^2}{q^2}\right)\right|_{D=4+2n-2\e}
\bbuildrel{=\!=\!=\!=}_{\scs q^2 \to \infty}^{}
\nonumber 
\eea 
\bea
\bbuildrel{=\!=\!=\!=}_{\scs q^2 \to \infty}^{}
\sum_{r=0}^{\infty} 
\left[ \left( \f{m^2}{q^2} \right)^r A^{ren}(k_1,k_2+n,2+n,r)
   +   \left( \f{m^2}{q^2} \right)^{r+\e} B(k_1,k_2+n,2+n,r) \right],
\eea
where 
\be
A^{ren}(k_1,k_2+n,2+n,r) = A(k_1,k_2+n,2+n,r) - 
\f{1}{\e(n+1)!} \delta_{1k_1} \delta_{1k_2} \delta_{0r}
\ee
The coefficients $A(k_1,k_2,\omega,r)$ and $B(k_1,k_2,\omega,r)$ are
given explicitly in appendix~C.

	At this point, we are ready to write down the desired
generalization of eqn.~(\ref{fin.sum})
\bea \label{fin.sum.n}
I^{(3)(n)}_{n_1 n_2 n_3 n_4 n_5 0} &=&
\mbox{(finite terms)} \;\; + 
\nonumber \\ && 
+ \; \; \f{(n_4)_n (n_5)_n (2-2\e)_n}{2^n (1-\e)_n}
\f{1}{\e \Gamma(2-\e)} 
\sum_{r_1=0}^{2+n-n_3} \;\; \sum_{r_2=0}^{2+n-n_3-r_1} 
\newton{-n_3}{2+n-n_3-r_1-r_2} 
\times \nonumber \\ && \hspace{2.5cm} \times
\left\{ A^{ren}(n_1,n_5+n,2+n,r_1) A^{ren}(n_2,n_4+n,2+n,r_2) \right.
\nonumber \\ && \hspace{3cm}
+ \f{1}{3} B(n_1,n_5+n,2+n,r_1) B(n_2,n_4+n,2+n,r_2) 
\nonumber \\ && \hspace{3cm}
+ \f{1}{2} \left[ A^{ren}(n_1,n_5+n,2+n,r_1) B(n_2,n_4+n,2+n,r_2) \right.  
\nonumber \\ && \hspace{3cm}
\left. \left. + B(n_1,n_5+n,2+n,r_1) A^{ren}(n_2,n_4+n,2+n,r_2) \right] \right\}
\nonumber \\ && \nonumber \\ && 
+ \f{n!}{\e\;2^n (n+1)} \sum_{\rho=0}^{[n/2]} \sum_{i=0}^{\rho} 
\sum_{j=0}^{n-2\rho} \sum_{l=0}^{j} \sum_{k=0}^{n-\rho-j} 
\newton{\rho}{i} \newton{n-2\rho}{j} \newton{j}{l} \newton{n-\rho-j}{k} 
\times \nonumber \\ && \hspace{1.5cm} \times
\f{(-1)^{n+\rho+i+k+l}}{\rho! (n-2\rho)! (n+1-\rho-\e)_{\rho}} 
\left[ \delta_{1n_1} \delta_{1n_5} I^{(2)}_{(n_2-i-l)(n_4-j+l)(n_3-k)}  \right.
\nonumber \\ && \hspace{7cm}
\left. + \delta_{1n_2} \delta_{1n_4} I^{(2)}_{(n_1-i-l)(n_5-j+l)(n_3-k)} \right]
\nonumber \\ && \nonumber \\ && 
- \delta_{1n_1} \delta_{1n_2} \delta_{1n_4} \delta_{1n_5} 
 \f{(2-2\e)_n(2-\e)_n}{\e^2\;2^n (n+1)^2 (1-\e)_n} \f{\Gamma(n_3-n-2+\e)}{\Gamma(n_3)}.
\eea

	The above equation is the main result of the present paper. It
gives us pole parts of all the nontrivial scalar three-loop integrals
$I^{(3)}$, i.e. those which do not reduce to products of lower-loop
integrals. When $n=0$, it reduces to eqn.~(\ref{fin.sum}).

\newpage \noindent
{\bf 5. From renormalization constants to $\beta$-functions and anomalous
dimensions.}

	In the preceding sections, we have described an algorithm for
calculating pole parts of Feynman diagrams. Using our formulae, one
can find all the MS-scheme renormalization constants in a given
theory, up to three loops. In the present short section, we give two
examples of relations between three-loop renormalization constants and
beta functions or anomalous dimensions.

	Here, we depart from the MS scheme and assume that the
renormalization constants (calculated in the framework of dimensional
regularization) can contain arbitrary finite terms. However, we assume
that these finite terms are renormalization-scale independent.

	For instance, let us consider renormalization of the gauge
coupling $g$ in some Yang-Mills theory 
%
\be
g^{\scs BARE} = \mu^{\e} \; Z_g \; g,
\ee
where $\mu$ is the renormalization scale. The renormalization constant
$Z_g$ has the following expansion in powers of the renormalized
coupling $g$:
\be
Z_g = 1 + g^2 \left( \kappa^{01} + \f{\kappa^{11}}{\e} \right) 
        + g^4 \left( \kappa^{02} + \f{\kappa^{12}}{\e} + \f{\kappa^{22}}{\e^2} \right) 
        + g^6 \left( \kappa^{03} + \f{\kappa^{13}}{\e} + \f{\kappa^{23}}{\e^2} 
                      + \f{\kappa^{33}}{\e^3} \right) + ...
\ee

	Some coefficients in this expansion are given in terms of the
others, which follows from locality of UV-divergences
\bea
\kappa^{22} &=& \f{3}{2} (\kappa^{11})^2 \nonumber \\
\kappa^{33} &=& \f{5}{2} (\kappa^{11})^3 \\
\kappa^{23} &=& \f{11}{3} \kappa^{11} \kappa^{12} - \f{7}{2} \kappa^{01} (\kappa^{11})^2.
\nonumber \eea

	From scale-independence of $g^{\scs BARE}$ one can derive the
following expression for the $\beta$-function in terms of
$\kappa^{ij}$:
\be \label{beta}
\beta(g) \equiv \mu \f{dg}{d\mu} = 2 \kappa^{11} g^3 +
\left[ 4 \kappa^{12} - 12 \kappa^{01} \kappa^{11} \right] g^5
+ \left[ 6 \kappa^{13} - 22 \kappa^{11} \kappa^{02} - 22 \kappa^{01} \kappa^{12} 
+ 54 \kappa^{11} (\kappa^{01})^2 \right] g^7 + ...
\ee

	As another example, let us discuss the anomalous dimension
matrix of a set of (possibly dimensionful) couplings $C_i$ which
linearly mix under renormalization
\be
C_j^{\scs BARE} = \sum_i C_i Z_{ij} = \left( \overline{C}^T \hat{Z} \right)_j.
\ee
Let us assume, that the renormalization constant matrix $\hat{Z}$
depends on a single gauge coupling $g$. Then it reads
\be
\hat{Z} = 1 + g^2 \left( \hat{a}^{01} + \f{\hat{a}^{11}}{\e} \right) 
        + g^4 \left( \hat{a}^{02} + \f{\hat{a}^{12}}{\e} + \f{\hat{a}^{22}}{\e^2} \right) 
        + g^6 \left( \hat{a}^{03} + \f{\hat{a}^{13}}{\e} + \f{\hat{a}^{23}}{\e^2} 
                      + \f{\hat{a}^{33}}{\e^3} \right) + ...
\ee

	Some coefficients in the above  expansion are given in terms of the
others, which follows from locality of UV-divergences
\bea
\hat{a}^{22} &=& \f{1}{2}(\hat{a}^{11})^2 + \kappa^{11} \hat{a}^{11} 
\nonumber \\
\hat{a}^{33} &=& \f{1}{6} (\hat{a}^{11})^3 + \kappa^{11} (\hat{a}^{11})^2
                  + \f{4}{3} (\kappa^{11})^2 \hat{a}^{11} \\
\hat{a}^{23} &=& 
\f{1}{3} \hat{a}^{11} \hat{a}^{12} 
+\f{2}{3} \hat{a}^{12} \hat{a}^{11} 
-\f{1}{6} \hat{a}^{01} (\hat{a}^{11})^2
-\f{1}{3} \hat{a}^{11} \hat{a}^{01} \hat{a}^{11} 
-\f{1}{3} \kappa^{11} \hat{a}^{01} \hat{a}^{11} 
+\f{4}{3} \kappa^{11} \hat{a}^{12}
+ \left( \f{4}{3} \kappa^{12} - \f{8}{3} \kappa^{01} \kappa^{11} \right) \hat{a}^{11}.
\nonumber \eea

	Scale-independence of $\overline{C}^{\scs BARE}$ implies that
the renormalized couplings $C_i$ satisfy the following renormalization
group equations
\be
\mu \f{d}{d\mu} \overline{C} = \hat{\gamma}^T \overline{C},
\ee
where the anomalous dimension matrix $\hat{\gamma}$ has the following
expansion in powers of $g$
%
\bea \label{gamma}
\hat{\gamma} &=& 2 \hat{a}^{11} g^2 
+ g^4 \left[ 4 \hat{a}^{12} - 2 \hat{a}^{01}\hat{a}^{11}  - 2 \hat{a}^{11}\hat{a}^{01}  
         - 4 \kappa^{11} \hat{a}^{01} - 4 \kappa^{01} \hat{a}^{11} \right] 
+ g^6 \left[ 6 \hat{a}^{13} 
            -4 \hat{a}^{12} \hat{a}^{01} 
            -2 \hat{a}^{01} \hat{a}^{12}
\right. \nonumber \\ && 
            -4 \hat{a}^{02} \hat{a}^{11} 
            -2 \hat{a}^{11} \hat{a}^{02} 
            +2 \hat{a}^{01} \hat{a}^{11} \hat{a}^{01} 
            +2 \hat{a}^{11} (\hat{a}^{01})^2
            +2 (\hat{a}^{01})^2 \hat{a}^{11} 
            +4 \kappa^{11} (\hat{a}^{01})^2
            -8 \kappa^{11} \hat{a}^{02} 
\nonumber \\ && \left.
            +4 \kappa^{01} \hat{a}^{11} \hat{a}^{01}
            +4 \kappa^{01} \hat{a}^{01} \hat{a}^{11} 
            -8 \kappa^{01} \hat{a}^{12}
            -8 \kappa^{12} \hat{a}^{01} 
            -8 \kappa^{02} \hat{a}^{11} 
           +24 \kappa^{01} \kappa^{11} \hat{a}^{01}
           +12 (\kappa^{01})^2 \hat{a}^{11} \right] + ... \hspace{1cm}
\eea

	In the MS scheme, equations (\ref{beta}) and (\ref{gamma})
become much simpler
\bea
\beta(g) &=& 2 \kappa^{11} g^3 + 4 \kappa^{12} g^5 + 6 \kappa^{13} g^7 + ...\\
\hat{\gamma} &=& 2 \hat{a}^{11} g^2 + 4 \hat{a}^{12} g^4 + 6 \hat{a}^{13} g^6 + ...
\eea
However, using the pure MS scheme may not be possible in some
effective theories where so-called ``evanescent operators'' arise in
dimensional regularization. This is why the more general relations
(\ref{beta}) and (\ref{gamma}) have been presented here.

\newpage \noindent
{\bf 6. Summary.}

	We have described an algorithm for calculating UV-divergent
parts of arbitrary Feynman diagrams. A common mass parameter has been
used to perform the infrared rearrangement. Explicit formulae for all
the necessary scalar integrals up to three loops have been given.  

	The main idea in calculating nontrivial three-loop integrals
was considering some of their two-loop subintegrals and expanding them
at large external momenta. In the end, some details have been given on
relations between UV-divergences and $\beta$-functions or anomalous
dimensions.

\ \\ 
{\bf Acknowledgments}

M.~Misiak thanks for hospitality at the University of Zurich where
most of this research has been performed.  K.~Chetyrkin appreciates
the warm hospitality of the Theoretical group of the Max Planck
Institute in Munich where part of this work has been made.
He also would like to thank S.A.~Larin for useful correspondence
related to Ref.~\cite{RVL97}.

	This work has been partially supported by the German
Bundesministerium f{\"u}r Bildung and Forschung under the contract 06 TM
874 and DFG Project Li 519/2-2. K.~Chetyrkin has been partially
supported by INTAS under Contract INTAS-93-0744-ext. M.~Misiak has been
supported in part by Schweizerischer Nationalfonds, by the Polish
Committee for Scientific Research (under grant 2~P03B~180~09,
1995--97) and by the EC contract HCMP CT92004.

\ \\ 
{\bf Appendix A}

	This appendix is devoted to reduction of tensor integrals to
scalar ones. We are interested in Feynman integrands depending only on
loop momenta and the artificial mass parameter $m^2$. The integrals
arising at one, two and three loops have the following form (in the
euclidean metric):
\be
(T^{(1)k}_n)_{\mu_1...\mu_k} = m^{-{\scs D}-k + 2 n}\; \pi^{-\f{D}{2}}
\int \f{ d^{\scs D} p \;\;\; p_{\mu_1}...p_{\mu_k}}{[p^2 +m^2]^n},
\ee
\be
(T^{(2)kl}_{n_1 n_2 n_3})_{\mu_1...\mu_k\nu_1...\nu_l} = 
m^{-2 {\scs D}-k-l + 2 \sum n_i}\; \pi^{-{\scs D}}
\int \f{ d^{\scs D} p \; d^{\scs D} q \;\;\;\; p_{\mu_1}...p_{\mu_k} \; q_{\nu_1}...q_{\nu_l}
}{[p^2 +m^2]^{n_1} [q^2 +m^2]^{n_2} [(p-q)^2 +m^2]^{n_3}}, \vspace{0.2cm}
\ee
\bea
(T^{(3)klm}_{n_1 n_2 n_3 n_4 n_5 n_6})_{\mu_1...\mu_k\nu_1...\nu_l\rho_1...\rho_m} = 
m^{-3 {\scs D} -k-l-m + 2 \sum n_i}\; \pi^{-\f{3D}{2}} \times
\hspace{5cm} && \nonumber
\eea
\be
\times \int \f{ d^{\scs D} p \; d^{\scs D} q \; d^{\scs D} r \;\;\;\;
p_{\mu_1}...p_{\mu_k} \; q_{\nu_1}...q_{\nu_l} \; r_{\rho_1}...r_{\rho_m}
}{[p^2 +m^2]^{n_1} [q^2 +m^2]^{n_2} [r^2 +m^2]^{n_3}
[(q-r)^2 +m^2]^{n_4} [(r-p)^2 +m^2]^{n_5} [(p-q)^2 +m^2]^{n_6}}.
\ee
Such integrals are proportional to linear combinations of products of
metric tensors. For instance,
\be \label{tensor.example}
(T^{(2)22}_{n_1 n_2 n_3})_{\mu \nu \rho \sigma}
=   F_1 \;  g_{\mu\nu}  g_{\rho\sigma} 
  + F_2 \; ( g_{\mu\rho} g_{\nu\sigma} + g_{\mu\sigma} g_{\nu\rho} ).
\ee
The tensors $g_{\mu\rho} g_{\nu\sigma}$ and $g_{\mu\sigma}
g_{\nu\rho}$ are multiplied by the same coefficient $F_2$ in the above
equation, due to an obvious symmetry. In a computer algebra code, such
a symmetry can be verified by checking that contractions of the l.h.s
of eqn.~(\ref{tensor.example}) with $g_{\mu\rho} g_{\nu\sigma}$ and
$g_{\mu\sigma} g_{\nu\rho}$ are identical. 

	The equations for the coefficients $F_1$ and $F_2$ are found
by contracting the tensor integral with $g_{\mu\nu} g_{\rho\sigma}$
and $g_{\mu\rho} g_{\nu\sigma}$
\be \left\{ \begin{array}{rcl}
D^2 F_1 + 2 D F_2   &=& X_1 \\
D F_1 + D (D+1) F_2 &=& X_2 
\end{array} \right.
\ee
where
\bea
X_1 &=& m^{-2 {\scs D}-4 + 2 \sum n_i}\; \pi^{-{\scs D}}
\int \f{ d^{\scs D} p \; d^{\scs D} q \;\;\;\; p^2 q^2
}{[p^2 +m^2]^{n_1} [q^2 +m^2]^{n_2} [(p-q)^2 +m^2]^{n_3}}, \\
X_2 &=& m^{-2 {\scs D}-4 + 2 \sum n_i}\; \pi^{-{\scs D}}
\int \f{ d^{\scs D} p \; d^{\scs D} q \;\;\;\; (p \cdot q)^2
}{[p^2 +m^2]^{n_1} [q^2 +m^2]^{n_2} [(p-q)^2 +m^2]^{n_3}}.
\eea
Consequently,
\be
\left( \begin{array}{c} F_1 \\ F_2 \end{array} \right) =
\left( \begin{array}{cc} D^2 & 2 D \\ D & D (D+1) \end{array} \right)^{-1}
\left( \begin{array}{c} X_1 \\ X_2 \end{array} \right).
\ee
The above matrix inversion is most easily done perturbatively in $\e$,
after substituting $D = 4 - 2 \e$. This makes the computer program
much faster, which is important for more complicated tensor integrals
where larger matrices need to be inverted.

	The integrals like $X_1$ and $X_2$ are easily reduced to the
standard scalar integrals (\ref{int2}), with help of the identities
\bea
p^2 &=& (p^2 +m^2) - m^2,\\
q^2 &=& (q^2 +m^2) - m^2,\\
p \cdot q &=& \f{1}{2} \{ (p^2 + m^2) + (q^2 + m^2) - [ (p-q)^2  + m^2 ] - m^2 \}.
\eea
	Specific values for the indices $n_i$ can be substituted only
after all these operations are performed.

\ \\
{\bf Appendix B}

	In this appendix, we give explicit formulae for the trivial
integrals, i.e. for the two-loop integrals which reduce to products of
one-loop ones, and for the three-loop integrals which reduce to
products of one- and two-loop ones.

	When at least one of the indices of the two-loop
integral $I^{(2)}_{n_1 n_2 n_3}$ is nonpositive, the integral
reduces to a product of tensor one-loop integrals. However, a
simple expression for such an integral can be also obtained from
a general formula for a two-loop integral with one massless and
two massive lines \cite{BV84}
\bea \label{i2.red}
I^{(2)}_{n_1 n_2 n_3}
&\bbuildrel{=\!=\!=\!=}_{\scs n_3 \leq 0}^{}&
\sum_{k=0}^{-n_3} \newton{-n_3}{k} m^{-2{\scs D}+2(n_1+n_2+k)} \pi^{-{\scs D}}
\int \f{ d^{\scs D} q_1 \; d^{\scs D} q_2}{[q_1^2+m^2]^{n_1} [q_2^2+m^2]^{n_2}
[(q_1-q_2)^2]^{-k}} \nonumber \\ \ \nonumber \\
&=& \sum_{k=0}^{-n_3} \newton{-n_3}{k} \f{ \Gamma(\f{D}{2}+k) 
\Gamma(n_1-k-\f{D}{2}) \Gamma(n_2-k-\f{D}{2}) \Gamma(n_1+n_2-k-D)}{
\Gamma(n_1) \Gamma(n_2) \Gamma(\f{D}{2}) \Gamma(n_1+n_2-2k-D)} \hspace{1.5cm}
\eea
	
	In the above equation, we have assumed that the nonpositive
index is $n_3$. This could have been done without loss of generality,
because $I^{(2)}_{n_1 n_2 n_3}$ is totally symmetric under
permutations of its indices. Equation (\ref{i2.red}) implies that
$I^{(2)}_{n_1 n_2 n_3}$ vanishes when more than one of its indices is
nonpositive.

	Let us now turn to the three-loop integrals $I^{(3)}_{n_1 n_2
n_3 n_4 n_5 n_6}$ considered in the cases B.1 and C.1 in the main
text. They have nonpositive index $n_6$ and, in addition, there is one
more nonpositive index among $n_1,n_2,n_4$ and $n_5$. Symmetries of
the diagram shown in fig.~\ref{double.bubble} allow to assume without
loss of generality that the other nonpositive index is $n_5$. The
remaining indices can be arbitrary integers. In such a case, the three
loop integral is expressible in terms of tensor one- and two-loop
integrals as follows:
\bea \label{reducible}
I^{(3)}_{n_1 n_2 n_3 n_4 n_5 n_6} =
\sum_{j_5=0}^{-n_5} \sum_{j_6=0}^{-n_6} \sum_{i_5=0}^{j_5} \sum_{i_6=0}^{j_6}
\sum_{k_5=0}^{-n_5-j_5} \sum_{k_6=0}^{-n_6-j_6} (-1)^{j_5+j_6} \; 2^{i_5+i_6}
\newton{-n_5}{j_5}\newton{-n_6}{j_6}\newton{j_5}{i_5}\newton{j_6}{i_6} \times
\nonumber
\eea
\be
\times \newton{-n_5-j_5}{k_5}\newton{-n_6-j_6}{k_6} 
\left( T^{(1)(i_5+i_6)}_{n_1-k_5-k_6} \right)^{\mu_1...\mu_{i_5+i_6}}
\left( T^{(2)i_6i_5}_{(n_2+n_6+j_6+k_6)(n_3+n_5+j_5+k_5)n_4} \right)_{\mu_1...\mu_{i_5+i_6}}
\ee
The tensor integrals appearing in the above equation have been defined
in appendix A.

	In the end, we consider the integral $I^{(3)(n)}_{n_1 n_2 n_3
n_4 n_5 0}$ defined in eqn.~(\ref{prefactor}) in the reducible case,
i.e. when $n_1, n_2, n_4$ or $n_5$ is nonpositive. In such a case, we
can calculate $I^{(3)(n)}_{n_1 n_2 n_3 n_4 n_5 0}$ by expressing it as
a linear combination of the integrals considered in the previous
paragraph
\bea
I^{(3)(n)}_{n_1 n_2 n_3 n_4 n_5 0} &=&  \sum_{\rho=0}^{[n/2]} \sum_{k=0}^{n-2\rho}
\sum_{i=0}^k \sum_{l_1=0}^{\rho} \sum_{l_2=0}^{k-i+\rho} \newton{n-2\rho}{k} 
\newton{k}{i} \newton{\rho}{l_1} \newton{k-i+\rho}{l_2}
\times \nonumber \\ && \times
\f{(-1)^{\rho+n+i+l_1+l_2} n!}{(n+1-\rho-\e)_{\rho} 2^n \rho! (n-2\rho)!}
\; I^{(3)}_{(n_1-i-l_1) (n_2-l_2) n_3 n_4 n_5 (k+2\rho-n)}.
\eea
In some of the integrals $I^{(3)}$ on the r.h.s. of the above
equation, we may need to permute the first five indices using the
tetrahedron symmetry. The fifth index must become nonpositive before
eqn.~(\ref{reducible}) is applied.

\newpage
{\bf Appendix C}

	Here, we give explicit formulae for the coefficients
$a(k_1,k_2,r)$ and $b(k_1,k_2,r)$ in the expansion (\ref{G2.exp}) of
$G_2$ at large $q^2$. In section 4, we also need their generalizations
to $D=2\omega-2\e$ dimensional space, with arbitrary $\omega$.  Thus,
we write
\bea
a(k_1,k_2,r) &=& A(k_1,k_2,\omega=2,r)\\
b(k_1,k_2,r) &=& B(k_1,k_2,\omega=2,r).
\eea

	The above quantities are symmetric with respect to their
first two arguments. Moreover, we are only interested in positive
$k_1$ and $k_2$ in our application to three-loop integrals.
Consequently, knowing $A$ and $B$ for $1 \leq k_1 \leq k_2$ is
everything we need here.

	Using eqns.~(18) and (A.1) of ref.~\cite{BD92},\footnote{
Elementary identities $\Gamma(z) \Gamma(z+1/2) = 2^{1-2z} \Gamma(2z) \;$ 
and $\;\Gamma(z) \Gamma(1-z) =
\pi/\sin(\pi z)\;$ have been used to simplify these expressions.}
one finds (for $1 \leq k_1 \leq k_2$)
\be
A(k_1,k_2,\omega,r) = \left\{ \begin{array}{cc}
\vspace{0.2cm}
0 & \mbox{when $r < k_1$ or $\f{k_1+k_2}{2} \leq r < k_2$}\\
\vspace{0.2cm}
\f{(-1)^{r-k_1}(r-1)!(k_2-r-1)! \Gamma(k_1+k_2-r-\omega+\e)}{
(k_1-1)!(k_2-1)!(r-k_1)!(k_1+k_2-2r-1)!}
& \mbox{when $k_1 \leq r < \f{k_1+k_2}{2}$}\\
\f{2 (r-1)!(2r-k_1-k_2)! \Gamma(k_1+k_2-r-\omega+\e)}{
(k_1-1)!(k_2-1)!(r-k_1)!(r-k_2)!} & \mbox{when $r \ge k_2$}
\end{array} \right. 
\ee
\be
B(k_1,k_2,\omega,r) = \left\{ \begin{array}{cc}
\vspace{0.2cm}
0 & \mbox{when $r < k_1 +k_2 -\omega$}\\
\f{ (-1)^{r-k_1-k_2+\omega} \Gamma(k_1-r-\e)\Gamma(k_2-r-\e)\Gamma(r+\e)}{
(k_1-1)!(k_2-1)!(r-k_1-k_2+\omega)!\Gamma(k_1+k_2-2r-2\e)}
& \mbox{when $r \geq k_1 +k_2 -\omega$}
\end{array} \right. 
\ee

	The coefficients $A$ and $B$ often contain simple poles
in $\e$. For convergent integrals, these poles are usually
also present, but they cancel out in the final expression for
$G_2$. Thus, even when both $G_2$ in eqn.~(\ref{eq.double.bubble})
are convergent, one needs to carefully keep track of the 
${\cal O}(\e)$ parts in $A$ and $B$.
	
\ \\ 
{\bf Appendix D}

	Here, we present a useful relation between tensor and scalar
one-loop integrals in different numbers of dimensions. Let
$p^{(\alpha_1...\alpha_n)}$ denote the only symmetric and
traceless\footnote{
Tracelessness of a rank $n$ tensor means vanishing of any of its
contractions with $g_{\mu\nu}$.}
rank $n$ tensor which can be formed from a D-vector $p$
(see eg. \cite{K96})
\be
p^{(\mu_1...\mu_n)} = \hat{S} \sum_{\rho=0}^{[n/2]} \f{(-1)^{\rho} n!}{
4^{\rho} \rho! (n-2\rho)! (n+1-\rho-\e)_{\rho}} g^{\mu_1 \mu_2}...
g^{\mu_{2\rho-1}\mu_{2\rho}} p^{2\rho} p^{\mu_{2\rho+1}}...p^{\mu_n}.
\ee
Here, $\hat{S}$ stands for the operator which symmetrizes with respect
to all $n$ indices and multiplies the result by $1/n!$.

	The tensors $p^{(\alpha_1...\alpha_n)}$ occur in a
useful relation between one-loop tensor and scalar integrals in
different numbers of dimensions \cite{Dtrick}
\bea
&m^{-{\scs D} + 2 k_1 + 2 k_2}& \pi^{-D/2}
\int d^{\scs D} p \f{p^{(\alpha_1...\alpha_n)}}{[p^2+m^2]^{k_1} [(p-q)^2+m^2]^{k_2}} =
\nonumber \\
&=& (k_2)_n q^{(\alpha_1...\alpha_n)} m^{-{\scs D}- 2 n + 2 k_1 + 2 k_2} 
\pi^{-D/2-n} \int d^{D+2n} p \f{1}{[p^2+m^2]^{k_1} [(p-q)^2+m^2]^{k_2+n}}
\nonumber \\ &=& (k_2)_n q^{(\alpha_1...\alpha_n)} 
G_2\left.\left(k_1,k_2+n,\f{m^2}{q^2}\right)\right|_{D=4+2n-2\e}.
\eea

	The tensor one-loop integrals one finds in calculating
scalar three-loop integrals are {\em not} given in terms of
symmetric and traceless tensors. Instead, one encounters powers
of scalar products of various $D$-vectors. However, these latter
objects can be reversibly related to contractions of symmetric
and traceless tensors. Let us define
\be
(p \cdot q)^{(n)} = p^{(\alpha_1...\alpha_n)} q_{(\alpha_1...\alpha_n)}.
\ee
In eqns. (A.10) and (A.15) of ref.~\cite{CT81}, the following
relations have been given:\footnote{
There is a misprint in eqn. (A.15) of ref.~\cite{CT81} which we correct here.}
\bea
(p \cdot q)^{(n)} &=& \sum_{\rho=0}^{[n/2]}
\f{(-1)^{\rho}}{(n+1-\rho-\e)_{\rho}} \f{n!}{4^{\rho}\rho!(n-2\rho)!}
(p^2 q^2)^{\rho} (p \cdot q)^{n-2\rho}\\
(p \cdot q)^n &=& \sum_{\rho=0}^{[n/2]}
\f{1}{(n+2-2\rho-\e)_{\rho}} \f{n!}{4^{\rho}\rho!(n-2\rho)!}
(p^2 q^2)^{\rho} (p \cdot q)^{(n-2\rho)}
\eea
This appendix summarizes all the information on symmetric and
traceless tensors necessary for deriving our final expressions
(\ref{decomp.into.sym}) and (\ref{fin.sum.n}) for the three-loop
integrals in the most complex case C.2.

\newpage
\setlength {\baselineskip}{0.2in}

\end{document}